\documentstyle[12pt,emlines]{article}
\def\be{\begin{equation}}
\def\ee{\end{equation}}
\begin{document}
\title{
\begin{flushright}
{\small SMI-05-98 }
\end{flushright}
\vspace{2cm}
On Large $N$ Conformal Theories,\\
Field Theories in Anti-De Sitter Space and\\ Singletons}
\author{
I.Ya. Aref'eva  and I.V. Volovich\\
$~~~~$  \\
{\it Steklov Mathematical Institute, Russian Academy of Sciences}\\
{\it Gubkin St.8, GSP-1, 117966, Moscow, Russia}\\
arefeva, volovich@mi.ras.ru}
\date {$~$}
\maketitle
\begin {abstract}
It was proposed by Maldacena that the large $N$ limit of certain
conformal field theories can be described in terms of supergravity
on anti-De Sitter spaces ({\it AdS}). Recently, Gubser, Klebanov and
Polyakov and Witten have conjectured that the generating functional for
certain correlation functions in conformal field theory is given
by the classical supergravity action on {\it AdS}.
It was shown that the spectra of states of the two theories are
matched and the two-point correlation
function was studied. We consider a  model of scalar
field with self-interaction and compare the
three- and four-point correlation functions computed from a classical
action on {\it AdS} with the large $N$ limit of conformal theory.
An extension of Maldacena's proposal is discussed. We argue that the large
$N$ limit of certain conformal field theories in a
$p$-brane background can be described in terms of supergravity on the
corresponding background. We discuss also the large $N$ limit for the
Wilson loop and suggest that singletons which according to Flato and
Fronsdal are constituents of composite fields in spacetime should obey the
quantum Boltzmann statistics.
\end{abstract}

\newpage
\section{Introduction}

The 't Hooft large $N$ limit in QCD  where $N$ is the number of colours
enables us to understand qualitatively certain striking
phenomenological features of strong interactions \cite {tH,Wit1,Pol}.
To perform an analytical consideration one needs to compute the sum of
all planar diagrams. The summation of planar diagrams has been performed
only in low dimensional space-times \cite {BIPZ}.

It was suggested  \cite {Wit1} that
a master field which dominates the large $N$ limit exists.
There was an old problem in quantum field theory  how to construct
the master field for the large $N$ limit in QCD.
This problem has been discussed in many works, for a review see for
example \cite {Mig}.  The problem has been reconsidered  more
recently \cite {GG}-\cite {AZ} by using methods of non-commutative
(quantum) probability theory.  There are basically two
 types of correlators in matrix theories.  The first type includes trace
 of operators in different points (one has such correlator, for example in
the Wilson loop) and the second one includes the product of traces of the
composite local operators.  A construction of the master field for the
Wilson type correlation functions has been proposed in  \cite {AVM}.  It
was shown that the master field satisfies to standard equations of
relativistic field theory but it is quantized according to  the so called
quantum Boltzmann relations
$$a_i a^*_j=\delta_{ij}$$
where $a_i$ and $a^*_j$ are
annihilation and creation operators.
These operators have a  realization in the
free (Boltzmannian) Fock space.
Quantum field theories in Boltzmannian Fock space has been
considered in \cite {AZ}.

It was suggested
 \cite{Vol}  that the Boltzmannian Fock space describing
the large $N$ limit of gauge theory should contains states
describing black holes which obey the quantum Boltzmann statistics,
i.e. black hole can be represented as a Boltzmann  gas of branes.
This suggestion was based on \cite{AVM},  on the computation of the black
hole entropy \cite{SV} and on the idea \cite{Tow} about condensate of
D0-branes in the large $N$ limit for the matrix regularisation of
membrane.  In \cite {BFKS,LT,Min,BFK} it was shown how to compute entropy
of black hole by using the Boltzmann gas model in Matrix theory
\cite{BFSS}.

Recently an exciting new development in the study of
the large $N$ limit for matrix models has been performed.
It was proposed by Maldacena \cite{Mal} that the large $N$ limit of
certain
superconformal field theories can be described in terms of supergravity
on anti-De Sitter spaces ({\it AdS}), see \cite{FrFe}-\cite {Kleba} for
further developments. Earlier computations of correlators
in the world volume theories are performed in
\cite{olderkleb,kkleb,GKT}. More recently, Gubser, Klebanov and Polyakov
\cite{GKP} and Witten \cite{Wit2} have conjectured that the generating
functional for certain correlation functions in superconformal field
theory is given by the classical supergravity action on {\it AdS}.  It was
shown that the spectra of states of the two theories are matched and this
conjecture was tested for the two-point correlation function.

In this note we discuss the interacting model of the scalar matrix field
and compare the three- and four-point correlation functions computed from
a classical action on {\it AdS} with the large $N$ limit of conformal
theory.
An extension of Maldacena's proposal is discussed. We argue that the large
$N$ limit of certain conformal field theories in a
$p$-brane background can be described in terms of supergravity on the
corresponding background.
We discuss also the large $N$ limit for the Wilson loop and
suggest that singletons which according to Flato and Fronsdal \cite{FlFr}
are constituents of composite fields in spacetime should obey the quantum
Boltzmann statistics.

\section{Conformal Field Theory}
We consider a field $\phi (x)$ in the Euclidean space $R^d$.
If one has a transformation law

\be
  \label {2.1}
\phi (\lambda x)\to\lambda ^{-\Delta}\phi(x)
\ee
under the scale transformation then the number $\Delta$ is called
the (scale)  dimension of the field $\phi$. If the value $\Delta$
is canonical then the two-point function is proportional
to the free zero mass propagator.  To construct
a non-trivial conformal invariant field theory we have to assume
that at least some of the fields have anomalous dimensions.
For a review of conformal field theory see for example \cite {TMP}.
If we have a set of fields $\phi_n(x)$ with dimensions $\Delta_n$
which transform under the infinitesimal conformal transformation

\be
   \label {2.2}
\delta x^{\mu}=x^{\mu} (\epsilon x)-\frac{1}{2}
\epsilon^{\mu}x^2
\ee
as

\be
     \label {2.3}
\delta \phi_n (x)=-\Delta_n(\epsilon x)\phi_n(x),
\ee
then one can derive the two, three and four-point correlation
functions in the following known form

\be
\label{2p}
<\phi (x_1) \phi (x_2)> =
\frac{C}{x_{12}^{2\Delta}},
\ee

\be
\label{3p}
<\phi_1 (x_1)\phi_2
(x_2)\phi_3 (x_3)> =
\frac{C_{123}}{x_{12}^{\Delta_1+\Delta_2-\Delta_3}
x_{23}^{\Delta_2+\Delta_3
-\Delta_1}x_{31}^{\Delta_3+\Delta_1-\Delta_2}},
\ee

\be
\label{4p}
<\phi_1 (x_1)\phi_2 (x_2)\phi_3 (x_3)\phi_4(x_4)> =
C_{1234}\prod_{ij}
x_{ij}^{\frac{1}{3}\Delta-\Delta_i-\Delta_j}f(\xi,\eta), \ee
where $\Delta=\Delta_1 +\Delta_2 +\Delta_3$ and

$$x_{ij}=|x_i-x_j|, ~~\xi=x_{12}x_{34}/x_{13}x_{24},
~~\eta=x_{12}x_{34}/x_{14}x_{23}.$$
Here $f(\xi,\eta)$ is an arbitrary function. So, conformal invariance
defines the two and three-point function up to a constant
and the four-point correlation function up to an
arbitrary function of two variables.  The problem is
how to fix the constants $C$'s and the function $f$
by using the known dynamics of the theory.

\section{The Large $N$ Limit of Matrix Theories}

Let us consider the model of an Hermitian scalar matrix
field $M(x)=(M_{ij}(x))$,~~ $i,j=1,2..N$
in the $d$-dimensional space $R^{d}$
with the action

\be
S=N \int d^{d}x[\frac{1}{2}Tr (\bigtriangledown M)^2 +\lambda Tr M^n].
\ee
We are interested in the computation of the large $N$ limit
for the following local correlation functions

\be
\frac{1}{N}<{\cal O}_{i_1}(x_1).....{\cal O}_{i_k}(x_k)>,
\label{3.1}
\ee
where

$$
{\cal O}_i(x)=Tr M^{i}(x)
$$
as well as the nonlocal
(Wilson's type) correlation functions

\be
\frac{1}{N}<{\cal O}_k (x_1,..x_k)>,
\label{3.3}
\ee
where

\be
{\cal O}_k (x_1,..x_k)=Tr(M(x_1)...M(x_k)).
\label{3.4}
\ee
Certainly they are related due to
\be
\label{rel}
:{\cal O}_k (x_1,..x_k):|_{x_i=x}={\cal O}_k (x).
\ee

In \cite{AVM} it was shown that the large $N$ limit of the Wightman
functions
of the form (\ref{3.3}) is governed by the Boltzmann master field.
In particular, for the Yang-Mills field $A_{\mu}(x)$ one has

\be
\lim _{N\to \infty}\frac{1}{N}<0|Tr {A_{\mu_1}(x_1)....A_{\mu_k}(x_k)}|0>
=(\Omega_0|B_{\mu_1}(x_1)...B_{\mu_k}(x_k)|\Omega_0).
\ee
Here $B_{\mu}(x)$ satisfies the Yang-Mills equation,
but it is quantized according to the Boltzmann commutation relations.
For the Wilson loop one has

\be
W(C)=\lim_{N \to \infty}\frac{1}{N}
<0|Tr Pexp \int_{C}A_{\mu}dx^{\mu}|0 >=
(\Omega_0|Pexp \int_{C}B_{\mu} dx^{\mu}|\Omega_0).
\ee

Now we consider a naive but perhaps illuminating extension of the
conjecture from \cite{GKP,Wit2} to the case of the scalar matrix
model (\ref{3.1}).
One conjectures the following representation for the generating functional
of the correlation functions in the large $N$ limit

\be
\frac{1}{N}<exp\{\int_{R^d}dx\Phi_0(x){\cal O}(x)\}>=e^{-I(\Phi)}
\label{3.41}
\ee
where ${\cal O}(x)=Tr M^{2d/(d-2)}(x)$.
Here $\Phi$ is a field in a $d+1$-dimensional space
$B_{d+1}$ such that $R^d$ is its boundary, $\partial B_{d+1}=R^d$.  The
functional $I(\Phi)$ in (\ref{3.41}) is equal to the value of the action
for the field $\Phi$ computed on the solution of the corresponding
equations of motion with the fixed value $\Phi_0$ on the boundary (i.e. on
$R^d$). We assume that the solution $\Phi$ is uniquely defined
by the boundary function $\Phi_0$. Therefore $I(\Phi)$ is in fact
a functional of $\Phi_0$. In the next section we consider an example
of computation of such a functional.

\section{The Dirichlet  Problem for the Non-linear Laplace Equation}

General problems of computation of  $n$-point correlation
functions in the boundary theory are discussed in \cite{Wit2}.
Here we carry out an explicit perturbative computation for a model
of scalar field with self-interaction. Hopefully this will help
to perform more complicated computations in supergravity.
The action for the conformal invariant scalar field is

\be
I=\int d^dx\sqrt{g}[-\Phi\Delta\Phi -\frac{d-2}{4(d-1)}R\Phi^2
+\lambda\Phi^{2d/(d-2)}]
\ee
We start with the discussion of the non-linear Laplace equation
in the flat space.
Let $\Omega$ be an open domain  in $R^{d+1}$ and consider the
Dirichlet problem

\be
\label{DP}
\Delta \Phi =\lambda \Phi ^{n-1} , ~~~ x\in \Omega\\
\ee

\be
\label{BC}
\Phi |_{\partial \Omega}=\Phi _{0}
\ee
in the flat metric. Using the Green function $G(x,y)$
satisfying

\be
\Delta G(x,y)=-\delta(x-y),
\ee

\be
G(x,y)|_{x\in \partial \Omega}=0
\ee
the solution of the problem
(\ref{DP}), (\ref{BC})
can be represented as the solution of the following integral equation
\cite{Vla}

\be
\label{IR}
\Phi (x)=-\int _{\partial \Omega}\frac{\partial G(x,y)}{\partial n_y}
\Phi _{0}(y)dS_{y} -\lambda \int _{\Omega} G(x,y)\Phi ^{n-1}(y)dy
\ee
One can get an expression for $\Phi(x)$ as a functional of $\Phi_{0}$
by expanding (\ref{IR})  in the perturbation series.
Evaluation of the action functional
\be
\label{Act}
I(\Phi)=\int _{\Omega} dx \sqrt{g}[\frac{1}{2}(\nabla \Phi)^2
+\frac{\lambda}{n} \Phi ^{n} ]
\ee
can be performed using the representation (\ref{IR}).

Let us consider as a simple  example the case of upper half space with
the flat metric. We use the following notations. For coordinates
in $R^{d+1}$ we use  notations
$x=(x_0,{\bf x})$, ${\bf x}=(x_1,...x_d)$,
$x^*=(-x_0,{\bf x})$,  and the  upper half space is

$$
R^{d+1}_+=\{(x_0, {\bf x}) \in R^{d+1}|x_0 >0\}$$
We also denote

\be
\label{strn}
|x-y|= \sqrt{(x_0-y_0)^2+|{\bf x}-{\bf y}|^2},
~~~|{\bf x}|^2=\sum _{i=1}^{d}x_i^2,~~~
|x-{\bf y}|= \sqrt{x_0^2+|{\bf x}-{\bf y}|^2}
\ee

The Green function for the half space has the form

\be
\label{GF}
G(x,y)=\frac{1}{b_d}(\frac{1}{|x-y|^{d-1}}-\frac{1}{|x-y^*|^{d-1}})
\ee
Using this expression in (\ref{IR}) one gets

$$
\Phi (x_0,{\bf x}) =c_d\int \frac{x_0\Phi _0({\bf y})}{|x-{\bf y}|^{d+1}}
d{\bf y}+
$$

\be
\label{FO}
+\lambda \frac{(c_d)^{n-1}}{b_d}\int dy_0d{\bf y}
[\frac{1}{|x-y|^{d-1}}-\frac{1}{|x-y^*|^{d-1}}]y_0^{n-1}
\prod _{i=1}^{n-1}(\frac{\Phi _0({\bf y}^{(i)})d{\bf y}^{(i)}}
{|y-{\bf y}^{(i)}|^{d+1}})
+....
\ee
$c_{d}=\Gamma (\frac{d+1}{2})/\pi^{\frac{d+1}{2}}$.
After simple calculations one gets the following representation
for the functional $I(\Phi )$

\be
\label{GF1}
I(\Phi ) =a_d\int \frac{\Phi _0({\bf x})\Phi _0({\bf y})}
{|{\bf x}-{\bf y}|^{d+1}} d{\bf x}d{\bf y}+
\lambda k_d \int dx_0 d{\bf x} x_0^n
\prod _{i=1}^{n}(\frac{\Phi _0({\bf y}^{(i)})d{\bf y}^{(i)}}
{|x-{\bf y}^{(i)}|^{d+1}})
+....
\ee

The quadratic part includes only integral over the boundary
and higher order terms include integration over the bulk
(Fig.1).  It is interesting to compare these computations with analogous
expressions in the functional integral approach to S-matrix \cite{AFS,LD}.

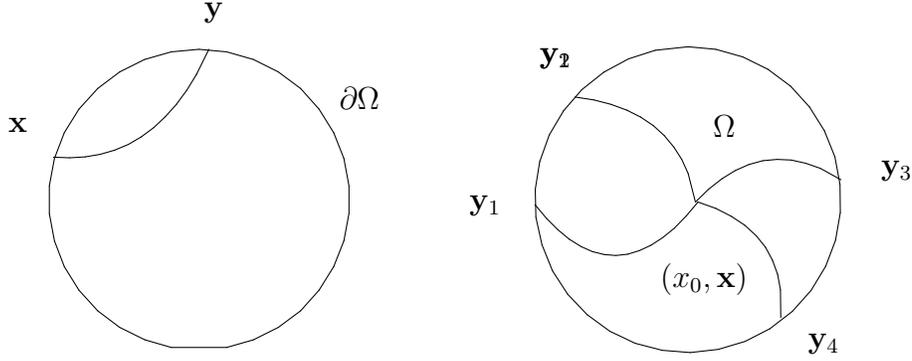
\begin{figure}
\begin{center}
\special{em:linewidth 0.4pt}
\unitlength 1.00mm
\linethickness{0.4pt}
\begin{picture}(125.67,64.33)
\emline{35.00}{60.01}{1}{38.68}{59.66}{2}
\emline{38.68}{59.66}{3}{42.23}{58.65}{4}
\emline{42.23}{58.65}{5}{45.53}{57.01}{6}
\emline{45.53}{57.01}{7}{48.48}{54.78}{8}
\emline{48.48}{54.78}{9}{50.96}{52.06}{10}
\emline{50.96}{52.06}{11}{52.91}{48.92}{12}
\emline{52.91}{48.92}{13}{54.24}{45.48}{14}
\emline{54.24}{45.48}{15}{54.92}{41.85}{16}
\emline{54.92}{41.85}{17}{54.92}{38.16}{18}
\emline{54.92}{38.16}{19}{54.24}{34.53}{20}
\emline{54.24}{34.53}{21}{52.91}{31.09}{22}
\emline{52.91}{31.09}{23}{50.97}{27.95}{24}
\emline{50.97}{27.95}{25}{48.48}{25.22}{26}
\emline{48.48}{25.22}{27}{45.53}{22.99}{28}
\emline{45.53}{22.99}{29}{42.23}{21.35}{30}
\emline{42.23}{21.35}{31}{38.68}{20.34}{32}
\emline{38.68}{20.34}{33}{31.33}{20.33}{34}
\emline{31.33}{20.33}{35}{27.78}{21.34}{36}
\emline{27.78}{21.34}{37}{24.47}{22.99}{38}
\emline{24.47}{22.99}{39}{21.53}{25.21}{40}
\emline{21.53}{25.21}{41}{19.04}{27.94}{42}
\emline{19.04}{27.94}{43}{17.10}{31.08}{44}
\emline{17.10}{31.08}{45}{15.76}{34.52}{46}
\emline{15.76}{34.52}{47}{15.08}{38.15}{48}
\emline{15.08}{38.15}{49}{15.08}{41.84}{50}
\emline{15.08}{41.84}{51}{15.76}{45.47}{52}
\emline{15.76}{45.47}{53}{17.09}{48.91}{54}
\emline{17.09}{48.91}{55}{19.03}{52.05}{56}
\emline{19.03}{52.05}{57}{21.52}{54.78}{58}
\emline{21.52}{54.78}{59}{24.46}{57.00}{60}
\emline{24.46}{57.00}{61}{27.77}{58.65}{62}
\emline{27.77}{58.65}{63}{31.32}{59.66}{64}
\emline{31.32}{59.66}{65}{35.00}{60.01}{66}
\emline{15.67}{45.67}{67}{17.83}{45.56}{68}
\emline{17.83}{45.56}{69}{19.90}{45.68}{70}
\emline{19.90}{45.68}{71}{21.89}{46.02}{72}
\emline{21.89}{46.02}{73}{23.79}{46.58}{74}
\emline{23.79}{46.58}{75}{25.60}{47.36}{76}
\emline{25.60}{47.36}{77}{27.33}{48.36}{78}
\emline{27.33}{48.36}{79}{28.97}{49.58}{80}
\emline{28.97}{49.58}{81}{30.52}{51.02}{82}
\emline{30.52}{51.02}{83}{31.99}{52.69}{84}
\emline{31.99}{52.69}{85}{33.37}{54.57}{86}
\emline{33.37}{54.57}{87}{34.66}{56.68}{88}
\emline{34.66}{56.68}{89}{36.33}{60.00}{90}
\emline{100.00}{60.34}{91}{103.72}{59.99}{92}
\emline{103.72}{59.99}{93}{107.31}{58.98}{94}
\emline{107.31}{58.98}{95}{110.66}{57.32}{96}
\emline{110.66}{57.32}{97}{113.65}{55.08}{98}
\emline{113.65}{55.08}{99}{116.17}{52.33}{100}
\emline{116.17}{52.33}{101}{118.15}{49.16}{102}
\emline{118.15}{49.16}{103}{119.52}{45.69}{104}
\emline{119.52}{45.69}{105}{120.23}{42.02}{106}
\emline{120.23}{42.02}{107}{120.26}{38.29}{108}
\emline{120.26}{38.29}{109}{119.61}{34.61}{110}
\emline{119.61}{34.61}{111}{118.29}{31.12}{112}
\emline{118.29}{31.12}{113}{116.36}{27.92}{114}
\emline{116.36}{27.92}{115}{113.88}{25.13}{116}
\emline{113.88}{25.13}{117}{110.92}{22.85}{118}
\emline{110.92}{22.85}{119}{107.60}{21.14}{120}
\emline{107.60}{21.14}{121}{104.03}{20.07}{122}
\emline{104.03}{20.07}{123}{100.31}{19.67}{124}
\emline{100.31}{19.67}{125}{96.59}{19.95}{126}
\emline{96.59}{19.95}{127}{92.98}{20.91}{128}
\emline{92.98}{20.91}{129}{89.61}{22.52}{130}
\emline{89.61}{22.52}{131}{86.59}{24.71}{132}
\emline{86.59}{24.71}{133}{84.02}{27.42}{134}
\emline{84.02}{27.42}{135}{81.99}{30.56}{136}
\emline{81.99}{30.56}{137}{80.57}{34.01}{138}
\emline{80.57}{34.01}{139}{79.80}{37.67}{140}
\emline{79.80}{37.67}{141}{79.71}{41.40}{142}
\emline{79.71}{41.40}{143}{80.31}{45.08}{144}
\emline{80.31}{45.08}{145}{81.57}{48.60}{146}
\emline{81.57}{48.60}{147}{83.46}{51.82}{148}
\emline{83.46}{51.82}{149}{85.90}{54.65}{150}
\emline{85.90}{54.65}{151}{88.81}{56.98}{152}
\emline{88.81}{56.98}{153}{92.11}{58.74}{154}
\emline{92.11}{58.74}{155}{95.67}{59.87}{156}
\emline{95.67}{59.87}{157}{100.00}{60.34}{158}
\emline{85.00}{53.67}{159}{87.49}{53.27}{160}
\emline{87.49}{53.27}{161}{89.77}{52.69}{162}
\emline{89.77}{52.69}{163}{91.85}{51.92}{164}
\emline{91.85}{51.92}{165}{93.72}{50.97}{166}
\emline{93.72}{50.97}{167}{95.39}{49.84}{168}
\emline{95.39}{49.84}{169}{96.85}{48.52}{170}
\emline{96.85}{48.52}{171}{98.11}{47.03}{172}
\emline{98.11}{47.03}{173}{99.16}{45.35}{174}
\emline{99.16}{45.35}{175}{100.00}{43.48}{176}
\emline{100.00}{43.48}{177}{101.00}{39.67}{178}
\emline{101.00}{39.67}{179}{102.74}{41.38}{180}
\emline{102.74}{41.38}{181}{104.51}{42.78}{182}
\emline{104.51}{42.78}{183}{106.30}{43.89}{184}
\emline{106.30}{43.89}{185}{108.12}{44.69}{186}
\emline{108.12}{44.69}{187}{109.96}{45.19}{188}
\emline{109.96}{45.19}{189}{111.83}{45.38}{190}
\emline{111.83}{45.38}{191}{113.72}{45.28}{192}
\emline{113.72}{45.28}{193}{115.63}{44.87}{194}
\emline{115.63}{44.87}{195}{117.57}{44.17}{196}
\emline{117.57}{44.17}{197}{120.33}{42.67}{198}
\emline{79.67}{39.33}{199}{81.19}{37.52}{200}
\emline{81.19}{37.52}{201}{82.72}{35.99}{202}
\emline{82.72}{35.99}{203}{84.25}{34.75}{204}
\emline{84.25}{34.75}{205}{85.79}{33.78}{206}
\emline{85.79}{33.78}{207}{87.33}{33.10}{208}
\emline{87.33}{33.10}{209}{88.87}{32.70}{210}
\emline{88.87}{32.70}{211}{90.42}{32.59}{212}
\emline{90.42}{32.59}{213}{91.96}{32.75}{214}
\emline{91.96}{32.75}{215}{93.52}{33.20}{216}
\emline{93.52}{33.20}{217}{95.07}{33.93}{218}
\emline{95.07}{33.93}{219}{96.63}{34.94}{220}
\emline{96.63}{34.94}{221}{98.19}{36.23}{222}
\emline{98.19}{36.23}{223}{99.76}{37.81}{224}
\emline{99.76}{37.81}{225}{101.33}{39.67}{226}
\emline{101.33}{39.67}{227}{103.69}{38.82}{228}
\emline{103.69}{38.82}{229}{105.77}{37.79}{230}
\emline{105.77}{37.79}{231}{107.56}{36.60}{232}
\emline{107.56}{36.60}{233}{109.07}{35.22}{234}
\emline{109.07}{35.22}{235}{110.30}{33.68}{236}
\emline{110.30}{33.68}{237}{111.25}{31.96}{238}
\emline{111.25}{31.96}{239}{111.92}{30.06}{240}
\emline{111.92}{30.06}{241}{112.30}{28.00}{242}
\emline{112.30}{28.00}{243}{112.33}{24.33}{244}
\put(53.67,51.99){$\partial \Omega$}
\put(103.33,48.33){$\Omega$}
\put(9.67,49.00){$\bf x$}
\put(35.67,64.33){$\bf y$}
\put(80.33,58.33){${\bf y}_1$}
\put(80.33,58.33){${\bf y}_2$}
\put(125.67,43.33){${\bf y}_3$}
\put(116.00,20.33){${\bf y}_4$}
\put(96.34,28.34){$(x_0,{\bf x})$}
\put(71.00,38.67){${\bf y}_1$}
\end{picture}

\end{center}
\caption{Two- and four-point correlators. Circles represent
the boundary $\partial \Omega$ of the domain $\Omega$. The four-point
correlator includes an integration over the bulk point $(x_0,{\bf
x})$}\label{fig1}
\end{figure}

Similar calculations performed for the upper half space  with
the Lobachevski metric

\be
ds^2= \frac{1}{x_{0}^2}dx^2
\ee
leads to the following  effective action

\be
\label{GFN}
I(\Phi ) =a_d\int \frac{\Phi _0({\bf x})\Phi _0({\bf y})}
{|{\bf x}-{\bf y}|^{2d}} d{\bf x}d{\bf y}+
\lambda k_d \int dx_0 d{\bf x} x_0^{d(n-1)-}
\prod _{i=1}^{n}(\frac{\Phi _0({\bf y}^{(i)})d{\bf y}^{(i)}}
{|x-{\bf y}^{(i)}|^{d}})
+....
\ee

Expressions (\ref{GF1}) and (\ref{GFN}) have been obtained by formal
manipulations and we ignored divergences.
We will present a thorough discussion of these
issues in another work. Here we will make only few remarks.
To make computations more rigorous it is convenient to use the Fourier
transform.

The solution of the Dirichlet problem
for the Laplace equation in the upper half space with the flat metric
\be
\label{fDp}
(\sum _{i=0}^{d}\frac{\partial ^2}{\partial x_{i}^{2}})\Phi=0,
~~~\Phi |_{x_0=0}=\Phi _{0}({\bf x})
\ee
can be represented in the form

\be
\label{ft}
\Phi (x_0,{\bf x})=c\int _{R^d}d{\bf p}e^{i{\bf px}-x_0|{\bf p}|}
{\tilde \Phi}_0 ({\bf p}),
\ee
where
\be
\label{fft}
{\tilde \Phi}_0 ({\bf p})=\int _{R^d}d{\bf x}e^{i{\bf px}}
 \Phi_0 ({\bf x}),
\ee
and we assume that $\Phi_0 ({\bf x})$ is a test function.
Then
\be
\label{dft}
\frac{\partial \Phi (x_0,{\bf x})}{\partial x_0}|_{x_0=0}=
-C\int _{R^d}d{\bf p}e^{i{\bf px}}|{\bf p}| {\tilde \Phi}_0 ({\bf p}),
\ee
and we get for the action
$$
I=\frac{1}{2} \int _{R_{+}^{d+1}}
 (\nabla \Phi_0 ({\bf x.x_0}))^2dx_0d{\bf x}=
\frac{1}{2} \int _{R^{d}}
d{\bf x} \Phi_0 \frac{\partial \Phi}{\partial x_0} =
$$
$$
-\frac{C}{2}\int _{R^{d}}d{\bf x} \Phi _0({\bf x})\cdot
\int _{R^d}d{\bf p}e^{i{\bf px}}|{\bf p}| {\tilde \Phi}_0 ({\bf p})=
$$
\be
\label{acft}
-\frac{C}{2}
\int _{R^d}d{\bf p}|{\bf p}| |{\tilde \Phi}_0 ({\bf p})|^2=
\int _{R^d}d{\bf x}\Phi ({\bf x}) \sqrt{-\Delta} \Phi_0 ({\bf x}).
\ee
All these formulae are well defined. $C$ denotes various constants.
Now
formally one can write the expression (\ref{acft})  as
\be
\label{for}
\int d{\bf x}d{\bf y} \Phi_0 ({\bf x})\frac{1}{|{\bf x}-{\bf y}|^{d+1}}
\Phi_0 ({\bf x}),
\ee
because
\be
\label{grf}
\int _{R^d} e^{i{\bf px}}|{\bf p}|d{\bf x}=
\frac{C}{|{\bf x}|^{d+1}}.
\ee
One interprets (\ref{for}) as the value of the distribution
$|{\bf x}-{\bf y}|^{-d-1}$ on a test function.
The distribution
$|{\bf x}|^{\lambda}$ is defined by means of the  analytical continuation
for $\lambda \neq -d, -d-2, -d-4,...$ \cite{GS}.

Now let us consider an AdS theory with action
\be
\label{Ads}
I=\frac{1}{2}\int _{\epsilon}^{\infty}dx_0 \int _{R^d}d{\bf x}
\frac{1}{x_{0}^{d-1}}\sum_{i=0}^{d}(\frac{\partial \phi}{\partial
x_{i}})^{2}
\ee
Here $\epsilon >0$ is a cut-off, see \cite{Polc,GKP}. Harmonic analysis
on AdS (Lobachevski-Poincare) spaces is considered
for example in \cite{Ahl,BR,Leng}.
The solution of the
Dirichlet problem
\be
\label{LDp}
(\sum _{i=0}^{d}\frac{\partial^2}{\partial x_{i}^{2}} -
\frac{(d-1)}{x_{0}}\frac{\partial}{\partial x_{0}}) \Phi=0,~~~
\Phi |_{x_0=0}=\Phi _{0}({\bf x})
\ee
can be represented
in the form
\be
\label{CD}
\Phi (x_0,{\bf x})=
cx_{0}^{d/2}\int _{R^d}d{\bf p}e^{i{\bf px}} |{\bf p}|^{\frac{d}{2}}
K_{\frac{d}{2}}(|{\bf p}|x_{0}){\tilde \Phi}_0 ({\bf p}),
\ee
where $K_{\frac{d}{2}}(y)$ is the
modified Bessel function.
By integrating by parts, one can rewrite
(\ref{Ads}) as

\be
\label{Adsa}
I=-\frac{1}{2}\int _{R^d}  d{\bf x}
(\frac{1}{x_{0}^{d-1}}\Phi \frac{\partial \Phi}{\partial x_0})
|_{x_0=\epsilon}
\ee
Using the asymptotic expansion  of the modified Bessel function
one gets a regularised expression for the action.

For $d=4$ one has  for $x_{0}\to 0$

\be
\label{CDA}
\Phi (x_0,{\bf x})=
C\int _{R^4}d{\bf p}e^{i{\bf px}}
[2-\frac{1}{2}(x_0p)^2-\frac{(x_0p)^4}{8}\ln\frac{x_0p}{2}+c(x_0p)^4+~...~]
{\tilde \Phi}_0({\bf p}),
\ee
here $p= |{\bf p}|$.
The action (\ref{Adsa}) for $\epsilon \to 0$ behaves as

\be
\label{bac}
I=
C\int _{R^4}d{\bf p}|{\tilde \Phi}_0({\bf p})|^2
[-\frac{1}{\epsilon ^2}p^2-\frac{p^4}{2}
\ln\frac{\epsilon p}{2}+c_1p^4+~...~].
\ee

 The appearance of divergent
terms in the classical action  can be  related with
the fact that the propagator for a field ${\cal O}$
of conformal dimension 4
should be a multiple of $|{\bf x}-{\bf y}|^{-8}$ and
one has to define it as a distribution.

In the spirit of the minimal subtractions scheme in the  theory
of renormalization one can write a "renormalized"
action as

\be
\label{rac}
I_{ren}=
C\int _{R^4}d{\bf p}|{\tilde \Phi}_0({\bf p})|^2
[-\frac{p^4}{2} \ln\frac{p}{2}+c_1p^4].
\ee

One can write the final result as follows
\be
\label{c}
I=\int _{R^5_+} dx\sqrt{g}(\nabla \Phi )^2
\longrightarrow
I_{ren}=    \int _{R^4}d{\bf x}
 \Phi _0\Delta ^2[c_1+c_2\ln (-\Delta)] \Phi _0
\ee
where the arrow includes the renormalization.
If one adds also  finite parts, then one gets  a term
$\Phi _0\Delta \Phi _0$.
One requires additional physical
assumptions  to fix the form of the renormalized action.

The renormalized action includes a local term
\be
\label{fac}
\int_{R^4}d{\bf x}(\Delta \Phi_0)^2
\ee
There is also a non-local term. This is related with the fact
that the distribution
$|{\bf x}|^{-8}$ \cite{GS} is not a homogeneous in $R^4$
(there is $log$ in the scaling law).

One can consider the renormalized
action  (\ref{rac}) as a definition of distribution $|\bf{x}-\bf{y}|^{-8}$
and  interpret the action \cite{Wit2}
\be
\label{wa}
\int_{R^{2d}} \frac{\Phi _0({\bf x})\Phi _0({\bf y})}
{|{\bf x}-{\bf y}|^{2d}} d{\bf x}d{\bf y}
\ee
as the value of the distribution $|{\bf x}-{\bf y}|^{-2d}$
on a test function.
The Fourier transform of this distribution \cite{GS}
includes a log-term. For $d=4$  one has
\be
\label{ftr}
\widetilde {|{\bf x }|^{-8}} =c_1p^4+c_2p^4\ln p,
\ee
which seems is in an agreement with \cite{GKP}.

\section{Discussion and Conclusion}

The supergravity solution in type IIB string theory carrying
D3-brane charge has the form
$$
ds^2=f^{-1/2}(-dt^2+dx_i^2)+f^{1/2}dy^2_{\mu}
$$
where $i=1,2,3,~\mu=4,5,6,7,8,9$ and $f=f(y)$ is a harmonic function.
If one has $N$ parallel D3-branes located at $y=0$ then
$$
f=1+\frac{4\pi gN\alpha'}{|y|^4}
$$
In the limit $gN>>1$ or $y\to 0$ one can neglect the 1 in the harmonic
function and the metric describes ${\it AdS}_5\times S^5$ space.
The world volume of $N$ parallel D3-branes is described
by ${\cal N}=4$ supersymmetric $U(N)$ gauge theory in 3+1 dimensions.
This is an essential point in the argument leading to the conjecture that
type IIB string theory on $({\it AdS}_5\times S^5)_N$ is dual to super
Yang-Mills theory \cite{Mal}.

Now let us consider two bunches of D3-branes located
at points $y^{(1)}$ and  $y^{(2)}$. In this case
$$
f= 1+4\pi g\alpha'[\frac{N_1}{|y-y^{(1)} |^4}
+\frac{N_2}{|y-y^{(2)} |^4}]
$$
There are  gravitational forces
between two bunches  and
it seems natural to think that this configuration
of branes is described by  super Yang-Mills theory
in the curved background. In the case $gN_1>>gN_2>>1$ one has super
Yang-Mills theory in the 3-brane background.  For a previous discussion
of M(atrix) theory in curved background along this line see
\cite{Vol2,ArV}. In the recent paper \cite{Kleba} there is
perhaps a related discussion of departures from conformal invariance.

It is known \cite{AVM} that the large N limit for correlation function of
composite operators
${\cal O}_k(x_1,...x_k)$ of the Wilson type (\ref{3.3}) is described by
the
Boltzmann quantum field theory.
Due to the relation (\ref{rel}) it seems natural to expect that singletons
which are constituents ("partons") of composite fields  in conformal
theory \cite{FlFr} also should obey the quantum Boltzmann statistics
in the large $N$ limit.

The expression (\ref{GFN}) obtained for correlation functions in the first
order of perturbation theory has an instructive form although
we didn't bring it to the conformal invariant form (\ref{3p}), (\ref{4p}).
Certainly we have discussed only the model with a simple interaction and
one has to look to the more complicated supergravity theory  including the
Kaluza-Klein modes to  check the conjecture \cite{GKP,Wit2}.

$$~$$
{\bf ACKNOWLEDGMENTS}
$$~$$

We are grateful to E.Witten for a helpful comment about
a relation between the asymptotic behavior of the classical action and
singularities in quantum correlators.
The work is supported in part by
INTAS grant 96-0698. I.A. is supported also by RFFI grant 96-01-00608 and
I.V. is supported  in part by  RFFI grant 96-01-00312.
 \end{document}